\documentclass[11pt,fleqn]{article} 
\usepackage{graphicx}
\usepackage{textcomp} 
\usepackage{epstopdf}
\usepackage{latexsym}
\usepackage{amsmath}
\usepackage{amssymb}
\usepackage{bm}
\usepackage[mathcal]{euscript}
\usepackage{slashed}
\usepackage{xcolor}

\usepackage[top=30mm, bottom=30mm, left=25mm, right=25mm, marginparwidth=12mm]{geometry}

\numberwithin{equation}{section}

\usepackage{marvosym}

\usepackage{indentfirst}

\newcommand\blfootnote[1]{
  \begingroup
  \renewcommand\thefootnote{}\footnote{#1}
  \addtocounter{footnote}{-1}
  \endgroup
}

\usepackage{hyperref}
\hypersetup{
        unicode,	
        colorlinks,
        citecolor=blue,
        linkcolor=blue, 
        urlcolor=red,
        bookmarksopen=true,
        bookmarksopenlevel=\maxdimen,
      }

\def\gl#1#2{\ifmmode \mathrm{GL}(#1; {\bf #2}) \else $\mathrm{GL}(#1; {\bf #2})$\fi}
\def\sl#1#2{\ifmmode \mathrm{SL}(#1; {\bf #2}) \else $\mathrm{SL}(#1; {\bf #2})$\fi}
\def\so#1{\ifmmode \mathrm{SO}({#1}) \else $\mathrm{SO}(#1)$\fi}

\def\sp#1#2{\ifmmode \mathrm{Sp}(#1; {\bf #2}) \else $\mathrm{Sp}(#1; {\bf #2})$\fi}
\def\usp#1{\ifmmode \mathrm{USp}(#1) \else $\mathrm{USp}(#1)$\fi}
\def\spin#1{\ifmmode \mathrm{Spin}(#1) \else $\mathrm{Spin}(#1)$\fi}
\def\su#1{\ifmmode \mathrm{SU}({#1}) \else $\mathrm{SU}(#1)$\fi}

\def\double #1{#1{\hbox{\kern-2pt $#1$}}}

\mathcode`\*="702A                  
\def\half{{\textstyle{1\over{\raise.1ex\hbox{$\scriptstyle{2}$}}}}}

\def \p{\partial}

\def \a{\alpha}

\def \b{\beta}

\def \d{\delta}

\def \g{\gamma}

\def \l{\lambda}

\def \o{\omega}

\def \O{\Omega}

\def \N{\nabla}

\def\ua{ {a}}
\def \r{\rho}

\def\ua{{ {a}}}
\def\ub{{ {b}}}
\def\uc{{ {c}}}

\begin{document}

\begin{flushright}
\makebox[0pt][b]{}
\end{flushright}

\vspace{40pt}
\begin{center}
{\LARGE A note on conserved worldsheet supercharges in heterotic pure spinor superstring}

\vspace{40pt}
Osvaldo Chandia${}^{\clubsuit}$ and Brenno Carlini Vallilo${}^{\spadesuit}$
\vspace{40pt}

{\em 
${}^{\clubsuit}$ Departamento de Ciencias, Facultad de Artes Liberales \\ Universidad Adolfo Ib\'a\~nez, Diagonal Las Torres 2640, Pe\~nalol\'en, Chile }\\

\vspace{20pt}

{\em 
${}^{\spadesuit}$ Departamento de F\'{\i}sica y Astronomía, Facultad de Ciencias Exactas\\ Universidad Andres Bello, Sazi\'e 2212, Santiago, Chile}

\vspace{60pt}
{\bf Abstract}
\end{center}

We study conserved worldsheet charges associated with spacetime supersymmetry in heterotic pure-spinor superstrings on curved ten-dimensional superspace backgrounds. Requiring BRST invariance and worldsheet conservation gives a covariant set of superspace constraints. In flat superspace these conditions reproduce the standard ten-dimensional supersymmetry generator. In curved superspace, they organize the requirements for global supersymmetry in terms of a normalizable spinor superfield.

\blfootnote{
${}^{\clubsuit}$ \href{mailto:ochandiaq@gmail.com}{ochandiaq@gmail.com},    
${}^{\spadesuit}$ \href{mailto:vallilo@unab.cl}{vallilo@unab.cl} }

\setcounter{page}0
\thispagestyle{empty}

\newpage

\tableofcontents

\parskip = 0.1in
\section{Introduction}
It is widely accepted that string theory or any theory of quantum gravity possesses no global symmetries \cite{Banks:1988yz,Banks:2010zn}. In the case of a quantum gravity theory with a holographic description, this was proved in \cite{Harlow:2018tng}. For string theory, the argument is that if there is a conserved worldsheet current, one can always construct a physical vertex operator for a spacetime field that will gauge the corresponding symmetry.  

Nevertheless, the conserved worldsheet currents and the associated conserved charges are useful for assessing the symmetries of a fixed string theory background. With this idea, we will study when a general four-dimensional supergravity background admits global supersymmetry. In past works \cite{Chandia:2022uyy,Chandia:2024rze}\footnote{See \cite{Chandia:2009it,Chandia:2011wd,Chandia:2011su} for earlier approaches for the specific cases of Calabi-Yau and $K3$ compactifications and \cite{Linch:2006ig,Linch:2008rw} for flux compactifications using the hybrid formalism \cite{Berkovits:1994wr}.} we have been exploring different approaches to describe compactifications directly from superspace. Although these works focus on superspace geometry and supergravity, the main motivation is their application to the pure spinor superstring \cite{Berkovits:2000fe}. Here we connect the ideas first presented in \cite{Chandia:2024rze} with the classical superstring by constructing the conserved charges associated with four-dimensional global supersymmetry in a fixed on-shell supergravity background. The key assumption in \cite{Chandia:2024rze} is the existence of a normalizable bosonic $SO(6)$ spinor superfield $\chi^I$ whose first component is the usual normalizable spinor in the internal manifold commonly used in the component approach to compactifications.

Although the motivation is four-dimensional supersymmetry after compactification, most of the derivation below is kept ten-dimensional and covariant. The four-dimensional interpretation enters through the choice of the spinor superfield $\chi^\alpha$, whose lowest component contains the internal spinor selecting the preserved supercharge. We therefore avoid an explicit component decomposition except when explaining the compactification interpretation. In the main derivation, Greek indices $\alpha,\beta,\ldots$ denote ten-dimensional Majorana-Weyl spinor indices. 

This paper is organized as follows. Section~\ref{yetanother} summarizes the heterotic pure-spinor action, BRST transformations, and background superspace constraints used below. Section~\ref{covcond} derives the covariant conditions imposed on a candidate conserved charge by BRST invariance and by worldsheet conservation. Section~\ref{constructcharge} constructs the charge in flat superspace and uses it as a consistency check of the general equations, before discussing the curved-superspace expansion relevant to compactifications. We close with conclusions and open questions.

\section{Pure spinor preliminaries}
\label{yetanother}

The basic variables of the heterotic pure-spinor string are the coordinates of a curved supermanifold ${\mathcal M}^{10|16}$, $Z^M$, a fermionic $SO(1,9)$ spinor $d_{ {\alpha}}$, a canonically conjugate pair of bosonic spinors $(\lambda^{ {\alpha}},\omega_{ {\alpha}})$, a set of right-moving heterotic fermions $\widetilde\psi$ and the right-moving $(\widetilde b,\widetilde c)$-ghosts. The heterotic fermions and right-moving ghosts will play no role in this work. The bosonic spinor satisfies the pure spinor condition
\begin{align}
    \lambda\gamma^a\lambda=0.
\end{align}

The heterotic pure-spinor action in a general supergravity background with trivial gauge fields is given by \cite{Berkovits:2001ue,Chandia:2003hn}
\begin{align}
S =\frac{1}{2\pi\alpha'} \int d^2z & \left( \frac12 \Pi_{  a} \widetilde\Pi^{  a} + \frac12 \Pi^A \widetilde\Pi^B B_{BA} + d_{ {\a}} \widetilde\Pi^{ {\a}} + \o_{ {\a}} \widetilde\N \l^{ {\a}} \right) + S_{\widetilde \psi} + S_{\widetilde b\widetilde c}+ S_{FT} ,
\label{actioncurved}
\end{align}
where $\Pi^A=\p Z^M E_M{}^A, \widetilde\Pi^A=\widetilde\p Z^M E_M{}^A$ where $E_M{}^A$ is the background supervielbein. The worldsheet covariant derivatives are defined as $\nabla=\Pi^A\nabla_A = \partial Z^M\nabla_M,\widetilde\nabla=\widetilde\Pi^A\nabla_A = \widetilde\partial Z^M\nabla_M$ where the background covariant derivative is defined as
\begin{align}
    \nabla_A=E_A{}^M\nabla_M= E_A{}^M\left(\partial_M+\frac{1}{2}\Omega_M{}^{\ua\ub}{\mathbf M}_{\ua\ub}+\Omega_M {\mathbf S}\right),
\end{align}
where ${\mathbf M}_{\ua\ub}$ and $\O_M{}^{\ua\ub}$ are the Lorentz generators and Lorentz connection, and $\O_M$ is the connection for the scaling generated by ${\mathbf S}$. This action has a symmetry generated by the pure spinor BRST charge
\begin{align}
    Q=\oint \l^{ {\a}}d_{ {\a}}.
\end{align}
Following \cite{Berkovits:2001ue}, the BRST charge is nilpotent and conserved whenever the background spacetime satisfies the ten-dimensional supergravity constraints. These constraints are
\begin{align}
    &T_{ {\a\b}}{}^{\ua}=-(\g^{\ua})_{ {\a\b}},\quad H_{ {\a\b}\ua}=-(\g_{\ua})_{ {\a} {\b}},\quad T_{\a ab}=2(\g_{ab}\O)_\a \cr
    &T_{ {\a} {\b}}{}^{ {\g}}=T_{\ua {\a}}{}^{ {\b}}=H_{ {\a\b\g}}=H_{\ua\ub {\a}}=H_{\ua\ub\uc}+T_{\ua\ub\uc}=0,
\end{align}
where $T$ denotes torsion components, $H$ denotes the components of $dB$, and $\g^{\ua}_{ {\a} {\b}}$ are the symmetric $16\times 16$ gamma-matrices in ten dimensions.
The curvature and torsion in superspace are defined by
\begin{align}\label{RandT}
    R_{A}{}^{B}=d\O_{A}{}^{B}+\O_{A}{}^{C}\O_{C}{}^{B},\quad T^{A}=\nabla E^A=dE^{A}+E^{B}\O_{B}{}^{A}.
\end{align}
Recall that these super-forms satisfy the Bianchi identities 
\begin{align}
    \nabla T^A = E^B R_B{}^A,
\end{align}
and the curvature two-form with vector and spinor indices are related by
\begin{align}
    R^{ab}=-\frac{1}{4}(\gamma^{ab})_\beta{}^\alpha R_\alpha{}^\beta.
\end{align}

The nilpotency of the BRST charge is verified by defining canonical variables in the action \eqref{actioncurved} and their canonical commutation relations,
\begin{align}
    [Z^M,P_N\}=\delta^M_N\delta(\sigma-\sigma'),\quad [\lambda^{ {\alpha}},\omega_{ {\beta}}]= \delta^{ {\alpha}}_{ {\beta}}\delta(\sigma-\sigma'),
\end{align}
where $(-1)^M P_M=\frac{\delta S}{\delta(\partial_\tau Z^M)}$. The conservation of the BRST charge is verified from the equations of motion derived from the action \eqref{actioncurved}. They are given by
\begin{align}
    &\widetilde\Pi^{ {\alpha}}=0,\quad 
    \widetilde\nabla\lambda^{ {\alpha}}=0,\quad 
    \widetilde\nabla\omega_{ {\alpha}}=0,\quad 
    \widetilde\nabla d_{ {\a}}=\lambda^{ {\beta}}\omega_{ {\gamma}}\widetilde\Pi^{\ua} R_{ {a\alpha\beta}}{}^{ {\gamma}},\\
    &\widetilde\nabla\Pi_\ua=d_{ {\alpha}} \widetilde\Pi^\ub T_{\ua\ub}{}^{ {\a}}+\lambda^{ {\alpha}}\omega_{ {\beta}}\widetilde\Pi^\ub R_{\ua\ub {\alpha}}{}^{ {\beta}},\\
    &\nabla\widetilde\Pi_\ua=-\Pi^\ub\widetilde\Pi^\uc H_{\ua\ub\uc}-\Pi^{ {\alpha}}\widetilde\Pi^\ub T_{\a\ua\ub}+d_{ {\alpha}}\widetilde\Pi^\ub T_{\ua\ub}{}^{ {\alpha}}
    +\lambda^{ {\alpha}}\omega_{ {\beta}}\widetilde\Pi^\ub R_{\ua\ub {\alpha}}{}^{ {\beta}},
\end{align}
where $R_{ABC}{}^D$ denotes the components of the curvature two-form. For an extensive list of the consequences derived from the supergravity constraints, nilpotency and conservation of the BRST charge, see \cite{Chandia:2022uyy}. In particular, the scale connection is 
\begin{align}
    \Omega_a=0, \quad \Omega_\alpha=\frac{1}{4}\nabla_\alpha\Phi, 
\end{align}
where $\Phi$ is the dilaton superfield. We will also use that \cite{Chandia:2024rze}
\begin{align}\label{DDphi}
    T_{\alpha a}{}^b&=2(\g_a{}^b\O)_\a,\cr
    \nabla_\alpha\nabla_\beta \Phi&= 
    \frac12\g^{ a}_{\a\b}\N_{ a} \Phi-\frac1{24}\g^{{abc}}_{\a\b} H_{{abc}},\\
    R_{\a\b ab}&=\N_{(\a}T_{\b)ab}+\g^c_{\a\b} H_{abc}\cr
    &=\frac12(\g_{[a})_{\a\b}\N_{b]}\Phi+\frac34\g^c_{\a\b}(H_{abc}-4(\O\g_{abc}\O))\cr &+\frac1{24}(\g_{abcde})_{\a\b}(H^{cde}-4(\O\g^{cde}\O)),\\
    R^{(S)}_{\alpha\beta}&=\frac{1}{4}\gamma^a_{\alpha\beta}\nabla_a\Phi,
\end{align}
where $R^{(S)}_{AB}$ is the scale curvature. The anti-commutator of spinor derivatives is then
\begin{align}\label{spinAntiComm}
    \{\nabla_\alpha,\nabla_\beta\}= \gamma^a_{\alpha\beta}\left(\nabla_a +\frac{1}{4}\nabla_a\Phi{\mathbf S}\right) +\frac{1}{2}R_{\alpha\beta}{}^{ab}{\mathbf M}_{ab}.
\end{align}

All these results are formulated in ten-dimensional curved spacetime. The purpose of this work is to study the consequences of manifest supersymmetry in lower dimensions, in particular in four spacetime dimensions. In Section \ref{comp} we will change the notation to better describe the decomposition to flat four-dimensional space and a compact six-dimensional manifold.

In the next section, we construct a generator for a Grassmann-odd global symmetry and identify the constraints imposed by BRST invariance in the pure spinor formalism.

\section{Covariant conditions for conserved supercharges}
\label{covcond}

The BRST transformations of the worldsheet fields generated by the BRST charge mentioned in Section \ref{yetanother} (up to a local Lorentz rotation \cite{Chandia:2006ix}) are given by
\begin{align}\label{Qws}
    &Q Z^M=\l^{ {\a}} E_{ {\a}}{}^M,\quad Q\o_{ {\a}}=d_{ {\a}},\quad Q\l^{ {\a}}=0,\cr 
    &Qd_{ {\a}}=-(\l\g_\ua)_\a\Pi^{\ua}+\l^{ {\b}}\l^{ {\g}}\o_{ {\r}} R_{ {\a\b\g}}{}^{ {\r}},
\end{align}
For the compactification discussion, one may decompose the ten-dimensional vector index as $a\to (a,i)$ and the ten-dimensional spinor index as $ {\a}\to(\a_I,\dot\a{}^I)$, using the conventions of \cite{Chandia:2024rze}. This will be used in Section \ref{comp}.

In deriving the constraints below, we assume that the background satisfies the standard heterotic pure-spinor superspace constraints. Equalities between integrated currents are understood modulo total derivatives on the worldsheet, and terms proportional to the pure-spinor constraint are set to zero. On dimensional grounds and in the purely classical limit,
the ansatz linear in \(d_\alpha\) and \(\Pi^A\) is then the most general one relevant for the spacetime supersymmetry charge.

In any background, the current $\chi^\alpha d_\alpha$, where $\chi^\alpha$ is a spinor superfield, is not separately BRST invariant or conserved. Additional terms proportional to $\Pi^A$ and $\omega_\alpha\lambda^\beta$ are required. In flat superspace these terms reduce to the familiar $\theta$-dependent pieces of the supersymmetry generator. We therefore consider the following covariant ansatz for the worldsheet current associated with a spacetime supersymmetry generator
\begin{align}\label{4dSUSY}
    q_\epsilon=\frac{1}{\alpha'}\oint\left( \chi^\alpha d_{\a}+\Pi^A f_{A}+\l^\a\o_\b g_\a{}^\b\right),
\end{align}
where $\chi^\alpha$,  $f_A$ and $g_\alpha{}^\beta$ are constrained by BRST invariance and conservation of the charge. We are using the label $\epsilon$ in $g_\epsilon$ to denote that $\chi^\alpha$,  $f_A$ and $g_\alpha{}^\beta$ depend on the parameter of the transformation generated by the charge. The pure spinor constraint implies there is a gauge transformation for the ghost conjugate momentum, $\delta\omega_\alpha= v_a (\lambda\gamma^a)_\alpha$, for any tangent vector $v_a$. Therefore, the charge above is consistent only if 
\begin{align}
    \lambda^\alpha (\lambda\gamma^a)_\beta g_\alpha{}^\beta=0.
\end{align}
This is solved by 
\begin{align}\label{gZeroFormTwoForm}
    g_\a{}^\b=\d_\a^\b g^{(0)}+ 
    \frac14 (\g^{ab})_\a{}^\b g^{(2)}_{ab}.  
\end{align}
Instead of imposing this form for $g_\alpha{}^\beta$, we will keep it general and show that BRST invariance and conservation also imply \eqref{gZeroFormTwoForm}.

Using the transformations in \eqref{Qws}, we obtain
\begin{align}\label{Qqa}
    Qq_\epsilon=\frac{1}{\alpha'}\oint \Big[&\l^{{\a}}\left(\N_{{\a}}\chi^\b+g_\a{}^\b\right) d_{\b}
    + \l^{ {\b}}\Pi^{ {\g}}\left(\g^{\ua}_{ {\b\g}}f_{\ua}-\N_{( {\b}}f_{ {\g})}\right)\cr
    &+\l^{ {\b}}\Pi^{\ua} \left(\N_{ {\b}}f_{\ua}-\N_{\ua}f_{ {\b}}+T_{ {\b}\ua}{}^{\ub}f_{\ub}+(\g_{\ua})_{ {\b}\alpha}\chi^\alpha \right)\cr
    &+\l^{ {\b}}\l^{ {\g}}\o_{ {\r}}\left(-R_{\a  {\b\g}}{}^{ {\r}}\chi^\alpha+\N_\b g_\g{}^\r \right) \Big],
\end{align}
where
\begin{align}\label{NchiNf}
    &\N \chi^\alpha=d\chi^\alpha+\chi^\beta\O_\beta{}^\alpha,\cr 
    &\N f_{A}=df_{A}-\O_{A}{}^{B} f_{B}.
\end{align}

The condition that \eqref{Qqa} vanishes implies 
\begin{align}
    &\N_\a\chi^\b+g_\a{}^\b=0,\label{Qq1}\\
    &\N_{(\a}f_{\b)}=\g^a_{\a\b}f_a,\label{Qq2}\\
    &\N_\a f_a-\N_a f_\a+T_{\a a}{}^bf_b=-(\g_a\chi)_\a,\label{Qq3}\\
    &\l^\a\l^\b\left(\N_\a g_\b{}^\g-\chi^\r R_{\r\a\b}{}^\g\right)=0.\label{Qq4}
\end{align}

Computing ${\dot q}$ and using the equation of motion $\widetilde\Pi^\a=0$, we obtain
\begin{align}
    \frac{\p q_\epsilon}{\p\tau}=&\frac{1}{\alpha'}\oint\left( d_\a\widetilde\Pi^a\left(-\N_a\chi^\a-T_{ab}{}^\a f^b\right)+\Pi^{ {a}}\widetilde\Pi^{ {b}}\left(\N_{ {b}}f_{ {a}}+T_{ {ba}}{}^{ {\b}}f_{ {\b}}\right)+\Pi^{ {\a}}\widetilde\Pi^{ {a}}\N_{ {a}}f_{ {\a}}\right.\cr
    &\left.+\l^{ {\a}}\o_{ {\b}}\widetilde\Pi^{ {a}}\left(\N_a g_\a{}^\b -R_{ {ab\a}}{}^{ {\b}}f^{ {b}}+\chi^\g R_{ {a}\g  {\a}}{}^{ {\b}}\right)\right),
\end{align} 
whose vanishing gives the equations
\begin{align}
    &\N_a\chi^\a+T_{ab}{}^\a f^b=0,\label{qp1}\\
    &\N_a f_\a=0,\label{qp2}\\
    &\N_b f_a+T_{ba}{}^\a f_\a=0,\label{qp3}\\
    &\N_a g_\a{}^\b+\chi^\g R_{a\g\a}{}^\b-R_{ab\a}{}^\b f^b=0.\label{qp4}
\end{align}

A conservative independent set of constraints is given by
\eqref{Qq1}, \eqref{Qq2}, \eqref{Qq3}, \eqref{qp1},
and \eqref{qp2}. The other equations are determined as follows.
Equation (\ref{Qq4}) is implied by (\ref{Qq1}), the Bianchi identity $R_{(\r\a\b)}{}^\g=0$ and the pure spinor condition. Equation (\ref{qp4}) is obtained as follows. Using (\ref{Qq1}) one obtains
\begin{align}
    \N_a g_\a{}^\b&=-\N_a\N_\a\chi^\b=[\N_\a,\N_a]\chi^\b-\N_\a\N_a\chi^\b\cr
    &=-T_{\a a}{}^b\N_b\chi^\b-\chi^\g R_{\a a\g}{}^\b-\N_\a\N_a\chi^\b,
\end{align}
we now express $R_{a\a\g}{}^\b$ as $R_{a(\a\g)}{}^\b-R_{a\g\a}{}^\b$ and use (\ref{qp1}) to get
\begin{align}
    \N_a g_\a{}^\b+\chi^\g R_{a\g\a}{}^\b=\chi^\g R_{a(\a\g)}{}^\b-T_{ab}{}^\b\N_\a f^b+\left(\N_\a T_{ab}{}^\b+T_{\a a}{}^c T_{cb}{}^\b\right)f^b,
\end{align}
finally, using the Bianchi identities $R_{a(\a\g)}{}^\b=\g^b_{\a\g}T_{ab}{}^\b$ and $\N_\a T_{ab}{}^\b=T_{\a[a}{}^cT_{b]c}{}^\b+R_{ab\a}{}^\b$ and use (\ref{Qq3}), (\ref{qp1}) and (\ref{qp2}) we obtain
\begin{align}
    \N_a g_\a{}^\b+\chi^\g R_{a\g\a}{}^\b-R_{ab\a}{}^\b f^b=0,
\end{align}
which corresponds to equation (\ref{qp4}). 

Equation (\ref{qp3}) is also implied. In fact, using (\ref{Qq2})
\begin{align}
    \g^a_{\a\b}\N_b f_a=\N_b\N_{(\a}f_{\b)}=[\N_b,\N_{(\a}]f_{\b)}+\N_{(\a}\N_b f_{\b)}=-R_{b(\a\b)}{}^\g f_\g,
\end{align}
where equation (\ref{qp2}) was used. Using the Bianchi identity $R_{b(\a\b)}{}^\g=\g^a_{\a\b}T_{ba}{}^\g$ we finally obtain (\ref{qp3}).

It is possible to find another expression for $g_\a{}^\b$ without the derivative of $\chi^\alpha$. From \eqref{Qq3} and \eqref{qp2} we have that
\begin{align}\label{alt1}
    -10\chi^\b=(\g^a)^{\b\g}\left(\N_\g f_a+18\O_\g f_a\right).
\end{align}
Applying $\N_\a$ to (\ref{alt1}) and using
 \begin{align}
    \nabla_\alpha\nabla_\gamma f_a = -\nabla_\gamma\nabla_\alpha f_a +(\gamma^b)_{\alpha\gamma}\nabla_b f_a + R_{\alpha\gamma ac} f^c,
\end{align}
we obtain
\begin{align}\label{alt2}
    10g_\a{}^\b+\g_a^{\b\g}\g^a_{\a\r} g_\g{}^\r&=\frac38(\g^{abc}\g^d)_\a{}^\b(\O\g_{abc}\O)f_d+\frac94(\g^a\g^b)_\a{}^\b(\N_a\Phi)f_b\cr
    &-\frac3{16}(\g^{abc}\g^d)_\a{}^\b T_{abc}f_d-(\g^{ab})_\a{}^\b T_{ab}{}^\g f_\g\cr 
    &+(\g^a)^{\b\g}\left(\N_\g T_{\a a}{}^b-R_{\a\g a}{}^b\right)f_b
    -(\g^a)^{\b\g}T_{\a a}{}^b\N_\g f_b\cr &+2(\g^a)^{\b\g}\O_\a\N_\g f_a
    -18(\g^a)^{\b\g}\O_\g\N_\a f_a+2\g_a^{\b\g}\g^a_{\a\r}\chi^\r\O_\g,
\end{align}
where we have used (\ref{DDphi}) and (\ref{qp3}). Using the Bianchi identity involving $R_{[\a\g a]}{}^b$ in the last term of the second line of (\ref{alt2}) and using (\ref{Qq3}) together with (\ref{qp3}) in terms with $\N_\a f_b$ in the last line of (\ref{alt2}) we finally get
\begin{align}\label{alt3}
    10g_\a{}^\b+\g_a^{\b\g}\g^a_{\a\r} g_\g{}^\r=&(\g^{ab})_\a{}^\b(-H_{ab}{}^c f_c-T_{ab}{}^\g f_\g)\cr
    &+20\chi^\b\O_\a-16\g_a^{\b\g}\g^a_{\a\r}\chi^\r\O_\g+2(\g^a\g^b)_\r{}^\b(\g_{ab})_\a{}^\g\chi^\r\O_\g.
\end{align}
Using the identity
\begin{align}\label{alt4}
    \chi^\b\O_\a=\frac1{16}\d_\a^\b\chi^\g\O_\g-\frac{1}{32}(\g^{ab})_\a{}^\b(\chi\g_{ab}\O)+\frac{1}{384}(\g^{abcd})_\a{}^\b(\chi\g_{abcd}\O),
\end{align}
equation (\ref{alt3}) becomes
\begin{align}\label{alt5}
    10g_\a{}^\b+\g_a^{\b\g}\g^a_{\a\r} g_\g{}^\r=-20\d_\a^\b\chi^\g\O_\g+(\g^{ab})_\a{}^\b(-H_{ab}{}^c f_c-T_{ab}{}^\g f_\g-2(\chi\g_{ab}\O)),
\end{align}
which implies
\begin{align}\label{gsolution}
    g_\a{}^\b=-\d_\a^\b \chi^\g\O_\g+\frac14(\g^{ab})_\a{}^\b(-H_{ab}{}^c f_c-T_{ab}{}^\g f_\g-2(\chi\g_{ab}\O)).
\end{align}
For later use, we will define 
\begin{align}\label{gcomponents}
    &g^{(0)}=-\chi^\alpha\O_\alpha, 
    \\
    &g^{(2)}_{ab}=-H_{ab}{}^c f_c-T_{ab}{}^\g f_\g-2(\chi\g_{ab}\O).
\end{align}

So far it seems that the independent set of equations is used to find $(f_a, \chi^\alpha, g_\alpha{}^\beta)$ in terms of $f_\alpha$, as in abelian SYM theory in ten dimensions. The main difference is that equation \eqref{qp2} was used to find the expression for $g_\alpha{}^\beta$. The remaining independent equation \eqref{qp2} is the one that restricts the class of on-shell superspaces admitting a global Grassmann odd symmetry. Since it contains the term $\nabla_a\chi^\alpha$, one immediately makes the connection to the covariant constancy associated with Killing spinors of supersymmetric backgrounds. However, as we will see later, it is useful to have an equation for $\chi^\alpha$ that does not involve its derivatives. Start with 
\begin{align}\label{lasteq}
    \{\N_\a,\N_\b\}\chi^\gamma +\nabla_\alpha g_\beta{}^\gamma+\nabla_\beta g_\alpha{}^\gamma=\g^a_{\a\b}\N_{a}\chi^\gamma+\chi^\sigma R_{\a\b\sigma}{}^\gamma +\nabla_\alpha g_\beta{}^\gamma+\nabla_\beta g_\alpha{}^\gamma=0,
\end{align}
Using \eqref{qp1} and the Bianchi identity $R_{(\alpha\beta\gamma)}{}^\delta=0$ we obtain
\begin{align}\label{Fundamental}
    -(\gamma^a)_{\alpha\beta}T_{ab}{}^\gamma f^b -
    \chi^\sigma R_{\sigma\alpha\beta}{}^\gamma
    -\chi^\sigma R_{\sigma\beta\alpha}{}^\gamma
    +\nabla_\alpha g_\beta{}^\gamma+\nabla_\beta g_\alpha{}^\gamma=0.
\end{align}
Using 
\begin{align}
    R_{\sigma\alpha\beta}{}^\gamma=\frac{1}{4}(\gamma^a)_{\sigma\alpha}\nabla_a\Phi\delta_\beta^\gamma +\frac{1}{4}(\gamma^{ab})_\beta{}^\gamma\left(\nabla_{(\sigma}T_{\alpha)ab}+(\gamma^c)_{\sigma\alpha} H_{abc} \right),
\end{align}
we can organize \eqref{Fundamental} as 
\begin{align}
    &-(\gamma^a)_{\alpha\beta}T_{ab}{}^\gamma f^b- \delta_\beta^\gamma\left(\frac{1}{4}(\chi\gamma^a)_\alpha \nabla_a\Phi -\nabla_\alpha g^{(0)} \right)\cr
    &+\frac{1}{4}(\gamma^{ab})_\beta{}^\gamma\left( \chi^\sigma \nabla_{(\sigma}T_{\alpha)ab}+(\chi\gamma^c)_{\alpha} H_{abc} +\nabla_\alpha g^{(2)}_{ab}\right)\cr
    &- \delta_\alpha^\gamma\left(\frac{1}{4}(\chi\gamma^a)_\beta \nabla_a\Phi -\nabla_\beta g^{(0)} \right)
    +\frac{1}{4}(\gamma^{ab})_\alpha{}^\gamma\left( \chi^\sigma \nabla_{(\sigma}T_{\beta)ab}+(\chi\gamma^c)_{\beta} H_{abc} +\nabla_\beta g^{(2)}_{ab}\right)=0.
\end{align} 
After contracting with $(\gamma_a)^{\alpha\beta}$, this equation becomes
\begin{align}\label{fundamental2}
     &-8T_{ab}{}^\gamma f^b-(\gamma_a)^{\gamma\alpha} 
    \left(\frac{1}{4}(\chi\gamma^b)_\alpha \nabla_b\Phi -\nabla_\alpha g^{(0)} \right)\cr 
    &+ 
    \frac{1}{4}(\gamma_a\gamma^{bc})^{\alpha\gamma} \left( -\chi^\sigma \nabla_{(\sigma}T_{\alpha)bc}-(\chi\gamma^d)_{\a} H_{bcd} +\nabla_\alpha g^{(2)}_{bc}\right)=0.
\end{align}
Note that $\nabla_\alpha g^{(0)}$ and $\nabla_\alpha g^{(2)}_{ab}$ contain $\nabla_\alpha \chi^\beta$, but they can be replaced using \eqref{Qq1}. We will discuss the consequence of this last equation in Section \ref{constructcharge}.

\section{Construction of the supercharge}
\label{constructcharge}

\subsection{Flat superspace}

As a consistency check, consider flat ten-dimensional superspace. All curvature and torsion components beyond $T_{\alpha\beta}{}^a=-(\gamma^a)_{\alpha\beta}$ vanish, so the general conditions for the existence of a Grassmann-odd conserved charge reduce to
\begin{align}
    &D_\alpha\chi^\beta+g_\alpha{}^\beta=0,\label{flat1}\\
    &D_{(\alpha}f_{\beta)}=\gamma^a_{\alpha\beta}f_a,\label{flat2}\\
    &D_{\alpha}f_{a}-\partial_a f_\alpha=-(\gamma_a\chi)_\alpha,\label{flat3}\\
    &\partial_a\chi^\alpha=0,\qquad
    \partial_a f_\alpha=0,\label{flat4}\\
    &\lambda^\alpha\lambda^\beta D_\alpha g_\beta{}^\gamma=0,\quad \partial_a g_\alpha{}^\beta=0\label{flat5}.
\end{align}
From \eqref{gsolution} one can immediately see that 
$g_\alpha{}^\beta=0$ in flat space.
Using this, equation \eqref{flat1}, together with \eqref{flat4}, implies that the spinor parameter is constant. We denote it by the explicit supersymmetry parameter
\begin{align}
    \chi^\alpha=\epsilon^\alpha .
\end{align}
The last equation in \eqref{flat4} implies that $f_a$ and $f_\alpha$ depend only on $\theta$.  Equation \eqref{flat3} is then solved by
\begin{align}
    f_a=-(\theta\gamma_a\epsilon).
\end{align}
Substituting this expression into \eqref{flat2} fixes the quadratic term in $f_\alpha$. With the ansatz
\begin{align}
    f_\alpha=A(\gamma_a\theta)_\alpha(\theta\gamma^a\epsilon),
\end{align}
one finds $A=1/3$, namely
\begin{align}\label{flatfalpha}
    f_\alpha=\frac{1}{3} (\gamma_a\theta)_\alpha(\theta\gamma^a\epsilon).
\end{align}
Therefore, the covariant equations derived above reproduce the standard flat-space supersymmetry charge,
\begin{align}
    q_\epsilon=\frac{1}{\alpha'}
    \oint\left[
    \epsilon^\alpha d_\alpha
    +(\epsilon\gamma_a\theta)\Pi^a
    +\frac{1}{3}(\gamma_a\theta)_\alpha(\theta\gamma^a\epsilon)\Pi^\alpha
    \right].
    \label{flatqepsilon}
\end{align}
This fixes the normalization of the compensating currents and provides the simplest check that the curved-background constraints have the correct flat limit.
Using the flat-space OPEs, this charge generates the standard supersymmetry transformations of flat superspace, up to normalization 
conventions for $d_\alpha$.

\subsection{Curved-superspace expansion}

Constructing the explicit component expansion of the conserved supercharge is not as simple as in the flat superspace case. 
The standard way of computing the component expansion in a curved superspace is to use normal-coordinate expansion in superspace \cite{McArthur:1983fm}. Around a point $Z_0$ one introduces tangent normal coordinates $Y^A$ and expands superfields covariantly along the corresponding superspace geodesic. For the present purpose it is enough to use fermionic normal coordinates, namely to expand only in the Grassmann-odd directions $Y^A=(0,\theta^\alpha)$ \cite{Gates:1997,Grisaru:2000}. The background dependence of the expansion is encoded in the normal-coordinate expansion of the background tensors, such as torsion, curvature, the $H$-field, dilaton, and their covariant derivatives. Once this background geometry is fixed, the only additional information needed to determine the coefficients in the expansions of $\chi^\alpha$, $f_a$ and $f_\alpha$ is the set of multiple Grassmann-odd covariant derivatives
\begin{align}
    &\nabla_{\alpha_1}\cdots \nabla_{\alpha_n} f_a\big|_{\theta=0},
    \qquad
    \nabla_{\alpha_1}\cdots \nabla_{\alpha_n} f_\beta\big|_{\theta=0},\\ \label{nablasChi}
    &\nabla_{\alpha_1}\cdots \nabla_{\alpha_n}\chi^\beta
    \big|_{\theta=0}=-\nabla_{\alpha_1}\cdots \nabla_{\alpha_{n-1}}g_{\alpha_n}{}^\beta\big|_{\theta=0},
\end{align}
which are obtained recursively from the constraint equations derived in Section \ref{covcond} and \eqref{gsolution}. 

The procedure is simplified by first considering dimensional analysis and a purely bosonic background at $\theta=0$. The supercharge $q_\epsilon$ should have mass dimension zero. With the conventions chosen in \eqref{actioncurved}, the worldsheet currents $(d_{ \alpha},\Pi^a,\Pi^{ \alpha},\omega_\alpha\lambda^\beta)$ have mass dimensions $(-3/2,-1,-1/2,-2)$. The supersymmetry parameter $\chi^\alpha$ has mass dimension $-1/2$, $f_{a}$ has mass dimension $-1$, $f_{ {\alpha}}$ has mass dimension $-3/2$ and $g_\alpha{}^\beta$ has mass dimension $0$. The lowest mass-dimensional background tensors that can appear in $q_\epsilon$ are $\nabla_a\Phi$ and $H_{abc}$. These considerations lead us to choose the lowest components of $\chi^\alpha$, $f_\alpha$,  $f_a$ and $g_\alpha{}^\beta$ to satisfy
\begin{align}\label{boundaryconditions}
    &\chi^\alpha\big|_{\theta=0}=\chi^\alpha_0,\cr
    &f_{ {\alpha}}\big|_{\theta=0}= 
    \nabla_{ \gamma} f_{ {\alpha}}\big|_{\theta=0}=0,\cr
    &f_{a}\big|_{\theta=0}=0,\cr
    &g_\alpha{}^\beta \big|_{\theta=0}=0.
\end{align}

Equations \eqref{Qq3} and \eqref{qp2} and the absence of Grassmann-odd background fields imply 
\begin{align}\label{ngfa}
    \nabla_{ \gamma} f_{a}\big|_{\theta=0}=- (\chi_0\gamma_a)_{\gamma}, 
\end{align}
Furthermore, equation \eqref{Qq2} implies that 
\begin{align}\label{solutionFalpha}
    \nabla_{ \beta}f_{ {\gamma}}=\frac{1}{2}\gamma^a_{ \beta \gamma} f_{a} + G_{ \beta \gamma},
\end{align}
where $G_{ \beta \gamma }$ is antisymmetric in $\beta$ and $\gamma$. Adding a total derivative to $q_\epsilon$ changes $f_A$ to $f_A+\nabla_A\Lambda$. Inserting this into \eqref{solutionFalpha}, we have 
\begin{align}
    \nabla_{ \beta}f_{ {\gamma}}+\nabla_\beta\nabla_\gamma\Lambda=\frac{1}{2}\gamma^a_{ \beta \gamma} f_{a}+\frac{1}{2} \gamma^a_{ \beta \gamma}\nabla_a\Lambda+ G_{ \beta \gamma}.
\end{align}
Using $\nabla_\beta\nabla_\gamma\Lambda=\frac{1}{2}[\nabla_\beta,\nabla_\gamma]\Lambda+\frac{1}{2}\gamma^a_{ \beta \gamma}\nabla_a\Lambda$, we obtain
\begin{align}
     \nabla_{ \beta}f_{ {\gamma}}+\frac{1}{2}[\nabla_\beta,\nabla_\gamma]\Lambda=\frac{1}{2}\gamma^a_{ \beta \gamma} f_{a} + G_{ \beta \gamma}.
\end{align}
Locally we can choose $\Lambda$ to eliminate $G_{\beta\gamma}$. Using this gauge choice, the second derivative of $f_\gamma$ can be computed
\begin{align}\label{secondDerivfalpha}
    \nabla_\alpha\nabla_\beta f_\gamma=
    -\frac{1}{4}\gamma^a_{ \beta \gamma}\nabla_\alpha\Phi f_{a}+
    \frac{1}{2}\gamma^a_{ \beta \gamma} \nabla_\alpha f_{a}. 
\end{align}
Since $\nabla_\alpha\Phi\big|_{\theta=0}=0$, 
\begin{align}\label{falphaSecondorder}
    \nabla_\alpha\nabla_\beta f_\gamma\big|_{\theta=0}= 
    -\frac{1}{2}\gamma^a_{ \beta \gamma}(\chi_0\gamma_a)_\alpha.
\end{align}

Up to this order in the expansion, we obtain exactly the flat space result. The second-order derivative of $f_\alpha$ in \eqref{falphaSecondorder} is the expected result from \eqref{flatfalpha} using the gamma matrix identity
\begin{align}
    (\gamma^a)_{\alpha\beta}(\gamma_a)_{\gamma\delta}+
    (\gamma^a)_{\beta\gamma}(\gamma_a)_{\alpha\delta}+ 
    (\gamma^a)_{\gamma\alpha}(\gamma_a)_{\beta\delta} =0.
\end{align}

The background tensors appear in the next non-trivial derivatives. The next non-vanishing derivatives of $\chi^\alpha$ can be computed from \eqref{Qq1} and \eqref{gsolution}:
\begin{align}
    \nabla_\alpha\nabla_\beta\chi^\gamma \big|_{\theta=0}&= -(\nabla_\alpha g_\beta{}^\gamma)\big|_{\theta=0} =\delta_\beta^\gamma\nabla_\alpha g^{(0)}\big|_{\theta=0} -\frac{1}{4}(\gamma^{ab})_\beta{}^\gamma \nabla_\alpha g^{(2)}_{ab}\big|_{\theta=0},
\end{align}
where
\begin{align}
    &\nabla_\alpha g^{(0)}\big|_{\theta=0}= -\frac{1}{4}\chi_0^\beta\nabla_\alpha\nabla_\beta\Phi\big|_{\theta=0},\\
    &\nabla_\alpha g^{(2)}_{ab}\big|_{\theta=0}= 
     H_{ab}{}^c\big|_{\theta=0}(\chi_0\gamma_c)_\alpha +  \frac{1}{2}\chi_0^\beta(\g_{ab})_\beta{}^\gamma 
     \nabla_\alpha\nabla_\gamma\Phi\big|_{\theta=0},
\end{align}
where $\nabla_\alpha\nabla_\beta\Phi$ is given by \eqref{DDphi}. 

Using again \eqref{Qq3}, \eqref{qp2}, \eqref{boundaryconditions} and the on-shell supergravity constraints we obtain for $f_a$
\begin{align}
    \nabla_\alpha\nabla_\beta\nabla_\gamma f_a\big|_{\theta=0} &=\frac12(\g_{abc})_{\alpha\beta}(\g^b\chi_0)_\gamma\N^c\Phi\big|_{\theta=0}-\frac1{48}(\g^{bcd})_{\alpha\beta}(\g_a\chi_0)_\gamma H_{bcd}\big|_{\theta=0}\cr 
     &-\frac18(\g_{abc})_{\alpha\beta}(\g_d\chi_0)_\gamma H^{bcd}\big|_{\theta=0}+\frac18(\g^{bcd})_{\alpha\beta}(\g_b\chi_0)_\gamma H_{acd}\big|_{\theta=0}\cr
     &-(\gamma_a)_{\gamma\delta} \nabla_\alpha\nabla_\beta\chi^\delta\big|_{\theta=0}.
\end{align}

Similarly, using \eqref{secondDerivfalpha} we obtain
\begin{align}
    \nabla_{\alpha_1}\nabla_{\alpha_2} \nabla_{\alpha_3}\nabla_{\alpha_4} f_\beta\big|_{\theta=0}&=
    -\frac{1}{4}\gamma^a_{\beta\alpha_4} (\nabla_{\alpha_2}\nabla_{\alpha_3}\Phi)\big|_{\theta=0}(\chi_0\gamma_a)_{\alpha_1}\cr 
    &+\frac{1}{4}\gamma^a_{\beta\alpha_4} (\nabla_{\alpha_1}\nabla_{\alpha_3}\Phi)\big|_{\theta=0} (\chi_0\gamma_a)_{\alpha_2}\cr
    &-\frac{1}{4}\gamma^a_{\beta\alpha_4}( \nabla_{\alpha_1}\nabla_{\alpha_2}\Phi)\big|_{\theta=0} (\chi_0\gamma_a)_{\alpha_3}
    +\frac{1}{2}\gamma^a_{\beta\alpha_4} 
    \nabla_{\alpha_1}\nabla_{\alpha_2} \nabla_{\alpha_3} f_a\big|_{\theta=0}. 
\end{align}
Higher-order terms are obtained recursively in the same way, although the expressions rapidly become lengthy. In the next order, there will be terms with $(\nabla\Phi)^2$, $\nabla\Phi H$, $H^2$, and the curvature.

To close this section, let us analyze the $|_{\theta=0}$ component of \eqref{fundamental2}. Since $T_{ab}{}^\alpha$ is Grassmann odd, $T_{ab}{}^\alpha|_{\theta=0}=0$.  Using \eqref{boundaryconditions} we obtain
\begin{align}\label{EQ1}
    &-(\gamma_a)^{\gamma\alpha} 
    \left(\frac{1}{4}(\chi_0\gamma^b)_\alpha \nabla_b\Phi\big|_{\theta=0}-\chi_0^\b\nabla_\alpha\O_\b\big|_{\theta=0}\right)\cr 
    &+ 
    \frac{1}{4}(\gamma_a\gamma^{bc})^{\alpha\gamma} \left(-\chi_0^\r\N_{(\r}T_{\a)bc}\big|_{\theta=0}+2(\chi_0\g_{bc})^\r\N_\a\O_\r\big|_{\theta=0} \right)=0,
\end{align}
recall that $T_{\a ab}=2(\g_{ab}\O)_\a$ and that
\begin{align}
    \N_\a\O_\b=\frac18\g^a_{\a\b}\N_a\Phi-\frac1{96}\g^{abc}_{\a\b}H_{abc},
\end{align}
we obtain that \eqref{EQ1} is equal to
\begin{align}\label{EQ2}
    &-\frac18\g_a^{\a\g}\chi_0^\b\left(\g^b_{\b\a}\N_b\Phi\big|_{\theta=0}-\frac1{12}\g^{bcd}_{\b\a}H_{bcd}\big|_{\theta=0}\right)\cr &-\frac1{16}(\g_{bc}\g_a\g^{bc})^{\a\g}\chi_0^\b\left(\g^b_{\b\a}\N_b\Phi\big|_{\theta=0}-\frac1{12}\g^{bcd}_{\b\a}H_{bcd}\big|_{\theta=0}\right)\cr 
    &=\frac{13}{4}\g_a^{\a\g}\chi_0^\b\left(\g^b_{\b\a}\N_b\Phi\big|_{\theta=0}-\frac1{12}\g^{bcd}_{\b\a}H_{bcd}\big|_{\theta=0}\right)=0,
\end{align}
which is proportional to the supersymmetry transformation of the dilatino. As we mentioned at the end of Section \ref{covcond}, the combined equations \eqref{Qq1} and \eqref{qp1} constrain the class of backgrounds admitting a global Grassmann odd symmetry. This final result strongly implies that the only possible symmetry of this type is indeed supersymmetry.

\subsection{Four dimensional compactification}
\label{comp}

The derivation above was kept ten-dimensional and covariant. We now briefly explain how the result is interpreted after compactification to four dimensions. By a $4+6$ compactification, we mean that, locally, the ten-dimensional target space is viewed as a fibration in which the non-compact directions describe four-dimensional spacetime. In contrast, the remaining six directions are internal. Correspondingly, tangent ten-dimensional vector indices are decomposed as $a=(\mu,i)$ where $\mu=0,\ldots,3$ labels the four-dimensional spacetime directions and
$i=1,\ldots,6$ labels the internal directions. We will maintain $a$ as the index for all bosonic directions. 

The ten-dimensional Majorana-Weyl spinor index is similarly decomposed under the corresponding subgroup of the Lorentz group, $SO(9,1)\rightarrow SO(3,1)\times SO(6).$ Schematically, we write $\alpha \rightarrow ({\alpha}_I,\dot\alpha{}^{I}),$ 
where $\alpha,\dot\alpha$ are now four-dimensional Weyl spinor indices and $I$ denotes an internal $SU(4)$ spinor index. Note that $(\alpha_I)^\dagger=(\dot\alpha^I)$. This notation is only meant to indicate the representation content. We do not need an explicit choice of $4+6$ gamma matrices for the derivation above.

The spinor superfield $\chi^\alpha$ then determines
which four-dimensional supersymmetry is generated by the worldsheet charge. For a background preserving a four-dimensional $N=1$ supersymmetry, one expects the corresponding ten-dimensional spinor parameter to factorize schematically as
\begin{align}\label{factorizingChi}
    \chi^\alpha \rightarrow \chi^{\alpha I}\oplus \bar\chi_{\dot\alpha I}=
    \epsilon^{\alpha}\,\eta^{I}(Z) \oplus
    \bar\epsilon_{\dot\alpha}\,\bar\eta_{I}(Z),
\end{align}
where $\epsilon^{\alpha}$ and $\bar\epsilon_{\dot\alpha}$ are the four-dimensional supersymmetry parameters 
while $\eta^{I}(Z)$ and $\bar\eta_I(Z)$ are internal spinors depending on the compact bosonic coordinates and all the Grassmann-odd coordinates. If $\eta^{I}(Z)$ and $\bar\eta_I(Z)$ are normalizable, i.e., $\eta^I\bar\eta_I\neq0$ globally, it is said that the internal manifold has $SU(3)$ structure. In this interpretation, the covariant equations obtained in Section \ref{covcond} are the worldsheet-current form of the supersymmetry-preservation conditions. Rather than imposing a component Killing-spinor equation from the supergravity point of view from the beginning, we characterize a preserved supersymmetry by the existence of superfields $(\eta^I,\bar\eta_I,f_{\alpha I},\bar f_{\dot\alpha}{}^I, f_\mu, f_i)$ satisfying the conditions derived in Section \ref{covcond}.

In the case of a compactification with four-dimensional Poincaré symmetry, we will assume
\begin{align}\label{assumptions}
    \nabla_\mu\Phi\big|_{\theta=0}= H_{\mu\bullet\bullet}\big|_{\theta=0} =R_{\mu\bullet}{}^{\bullet\bullet}\big|_{\theta=0} =R_{\bullet\bullet}{}^{\mu\bullet}\big|_{\theta=0}=0, 
\end{align}
where $\bullet$ means four-dimensional or six-dimensional tangent vector indices. These assumptions are imposed only on the lowest components. The higher components are constrained recursively by the superspace equations. Imposing the same conditions on the full superfields would be unnecessarily restrictive. The torsions $(T_{\mu\nu}{}^{\alpha I},T_{\mu i}{}^{\alpha I}, T_{\alpha I \mu}{}^\nu, T_{\alpha I \mu}{}^i)$ and curvature $R_{\mu \alpha I}{}^{\bullet\bullet}$, and their conjugates, are Grassmann odd; their $\theta=0$ components identically vanish in a purely bosonic background. 

Since we are using \eqref{factorizingChi}, the equation  \eqref{EQ2} gives two equations: one proportional to $\epsilon_\alpha$ and the other proportional to $\epsilon_{\dot\alpha}$. The first is 
\begin{align}
    \eta^I\left(\gamma^i_{IJ}\nabla_i\Phi -\frac{1}{12}(\gamma^{ijk})_{IJ}H_{ijk}\right)=0.
\end{align}
The $\theta=0$ projection of the left-hand side of the equation above is the supersymmetry transformation of the dilatino, which should vanish in supersymmetric backgrounds with torsion, as first analyzed in the seminal work \cite{Strominger:1986uh}. 

The superspace equation \eqref{qp1} at $\theta=0$ gives
\begin{align}
    \nabla_\mu (\epsilon\eta)\big|_{\theta=0}=0, \quad 
    \nabla_i (\epsilon\eta)\big|_{\theta=0}=0.
\end{align}
The first, given the assumptions \eqref{assumptions}, is just the constancy of the supersymmetry parameter in the four-dimensional spacetime. To interpret the second one, we must remember that the connection in $\nabla_i$ is torsionful, so it corresponds to the textbook local supersymmetry transformation for the gravitino. This condition only implies Ricci flatness if the dilaton is constant and $H=0$. Acting with $\nabla_{[\alpha}\nabla_{\beta]}$ on \eqref{qp1}, projecting to $\theta=0$ and using \eqref{boundaryconditions}, we obtain
\begin{align}
    \nabla_a\nabla_{[\alpha} g_{\beta]}{}^\gamma\big|_{\theta=0} +\chi^\sigma_0 \nabla_{[\alpha} R_{\beta]a\sigma}{}^\gamma \big|_{\theta=0} 
    +2(\chi_0\gamma^b)_{[\alpha} \nabla_{\beta]}T_{ab}{}^\gamma\big|_{\theta=0}=0,
\end{align}
which will impose an algebraic relation between $\chi$, $\nabla^2\Phi$, $(\nabla\Phi)^2$, $\nabla H$, $\nabla\Phi H$, $H^2$ and the curvature. 

The important point is that these equations were not imposed directly as component supersymmetry variations. They arise as projections of the covariant conditions required for the existence of a conserved worldsheet charge. The full superspace system determines the higher $\theta$-components of $\chi^\alpha$, together with the compensating superfields $f_A$ and $g_\alpha{}^\beta$, recursively in terms of the background torsion, curvature, flux and dilaton. A complete solution of this recursive system would give the full superspace lift of the four-dimensional preserved supercharge. In this paper we only need the lowest-component projection, which shows how the usual compactification conditions are encoded in the conserved-current construction.

\section{Conclusions and prospects}

In this note we formulated the existence of conserved worldsheet charges associated with four-dimensional supersymmetry directly in ten-dimensional heterotic superspace. Starting from the curved-background pure-spinor action, we imposed BRST invariance and worldsheet conservation on the general charge \eqref{4dSUSY}.  The resulting covariant system, summarized in \eqref{Qq1}--\eqref{Qq4} and \eqref{qp1}--\eqref{qp4}, gives the superspace conditions on the spinor superfield $\chi$ and on the compensating coefficients $f_A$ and $g_\alpha{}^\beta$.

The flat-superspace analysis provides a useful normalization and consistency check. The solution \eqref{flatqepsilon} reproduces the standard spacetime supersymmetry generator, with $\epsilon^\alpha$ denoting the constant supersymmetry parameter. For curved backgrounds, the same equations provide a covariant starting point for discussing four-dimensional global supersymmetry without introducing an explicit component decomposition in the main derivation. In the compactification interpretation, the lowest component of $\chi$ is naturally identified with the internal spinor that selects the preserved supersymmetry. In contrast, the higher $\theta$ components and the coefficients $f_A$ are determined recursively by the background torsion, curvature, flux, and dilaton data. A complete solution of this recursive system and its comparison with the usual component Killing-spinor equations are natural next steps.

Throughout the analysis, we have worked at the classical level in worldsheet theory; possible quantum corrections to current conservation are not addressed here. Another interesting direction is to include a gauge background and the heterotic fermions in the conserved charge. This will probably give a way to describe the heterotic string in $AdS_4$ with a gaugino condensate \cite{Frey:2005zz}. A more speculative project is to use $\chi$ and related superfields to construct worldsheet holomorphic currents to describe $N=2$ superconformal generators, which naturally appear in RNS Calabi-Yau compactifications. The methods used to construct the $b$ ghost in the pure spinor formalism 
\cite{Chandia:2013ima,Berkovits:2014ama,Chandia:2021coc} may help in this direction. Finally, we would like to comment about a new superstring formalism that mixes RNS, GS and pure spinors strings \cite{Berkovits:2021xwh}. The corresponding versions in curved backgrounds were developed in \cite{Berkovits:2022dbm}, \cite{Chandia:2023eel} and \cite{Chandia:2025nxx}. It would be interesting to adapt our methods to this new formalism.

\vskip 0.2in
{\bf Acknowledgments:} The work of BCV is partially supported by FONDECYT Regular grant 1250672.

{\bf Disclaimer:} During the preparation of this work, the authors utilized Generative AI for structural organization and language refinement. The authors take full responsibility for the manuscript's content, including all ideas, mathematical derivations, and conclusions.

\appendix

{
\bibliographystyle{abe}
\bibliography{mybib}{}

@article{Banks:2010zn,
    author = "Banks, Tom and Seiberg, Nathan",
    title = "{Symmetries and Strings in Field Theory and Gravity}",
    eprint = "1011.5120",
    archivePrefix = "arXiv",
    primaryClass = "hep-th",
    doi = "10.1103/PhysRevD.83.084019",
    journal = "Phys. Rev. D",
    volume = "83",
    pages = "084019",
    year = "2011"
}

@article{Banks:1988yz,
    author = "Banks, Tom and Dixon, Lance J.",
    title = "{Constraints on String Vacua with Space-Time Supersymmetry}",
    reportNumber = "PUPT-1086, SCIPP-8805",
    doi = "10.1016/0550-3213(88)90523-8",
    journal = "Nucl. Phys. B",
    volume = "307",
    pages = "93--108",
    year = "1988"
}

@article{Gates:1997,
  author        = {Gates, S. J. and Grisaru, Marcus T. and Knutt-Wehlau, Marcia E. and Siegel, Warren},
  title         = {Component Actions from Curved Superspace: Normal Coordinates and Ectoplasm},
  journal       = {Phys. Lett. B},
  volume        = {421},
  pages         = {203--210},
  year          = {1998},
  eprint        = {hep-th/9711151},
  archivePrefix = {arXiv}
}

@article{Grisaru:2000,
  author        = {Grisaru, Marcus T. and Knutt, Marcia E.},
  title         = {Norcor vs the Abominable Gauge Completion},
  journal       = {Phys. Lett. B},
  volume        = {500},
  pages         = {188--194},
  year          = {2001},
  eprint        = {hep-th/0011173},
  archivePrefix = {arXiv}
}

@article{Chandia:2024rze,
    author = "Chandia, Osvaldo and Vallilo, Brenno Carlini",
    title = "{Compactifications of Type II supergravities in superspace}",
    eprint = "2405.04736",
    archivePrefix = "arXiv",
    primaryClass = "hep-th",
    doi = "10.1007/JHEP11(2024)118",
    journal = "JHEP",
    volume = "11",
    pages = "118",
    year = "2024"
}

@article{Harlow:2018tng,
    author = "Harlow, Daniel and Ooguri, Hirosi",
    title = "{Symmetries in quantum field theory and quantum gravity}",
    eprint = "1810.05338",
    archivePrefix = "arXiv",
    primaryClass = "hep-th",
    doi = "10.1007/s00220-021-04040-y",
    journal = "Commun. Math. Phys.",
    volume = "383",
    number = "3",
    pages = "1669--1804",
    year = "2021"
}

@article{Chandia:2022uyy,
    author = "Chandia, Osvaldo and Vallilo, Brenno Carlini",
    title = "{Superspaces for heterotic pure spinor string compactifications}",
    eprint = "2205.01765",
    archivePrefix = "arXiv",
    primaryClass = "hep-th",
    doi = "10.1140/epjc/s10052-022-10947-0",
    journal = "Eur. Phys. J. C",
    volume = "82",
    number = "11",
    pages = "991",
    year = "2022"
}

@article{Strominger:1986uh,
    author = "Strominger, Andrew",
    title = "{Superstrings with Torsion}",
    reportNumber = "NSF-ITP-86-16",
    doi = "10.1016/0550-3213(86)90286-5",
    journal = "Nucl. Phys. B",
    volume = "274",
    pages = "253",
    year = "1986"
}

@article{Frey:2005zz,
    author = "Frey, Andrew R. and Lippert, Matthew",
    title = "{AdS strings with torsion: Non-complex heterotic compactifications}",
    eprint = "hep-th/0507202",
    archivePrefix = "arXiv",
    reportNumber = "CALT-68-2567, UK-05-05",
    doi = "10.1103/PhysRevD.72.126001",
    journal = "Phys. Rev. D",
    volume = "72",
    pages = "126001",
    year = "2005"
}

@article{Chandia:2011su,
	Archiveprefix = {arXiv},
	Author = {Chandia, Osvaldo and Linch, III, William D. and Vallilo, Brenno Carlini},
	Eprint = {1109.3200},
	Primaryclass = {hep-th},
	Slaccitation = {%%CITATION = ARXIV:1109.3200;%%},
	Title = {{The Covariant Superstring on K3}},
	Year = {2011}}

@article{Linch:2008rw,
	Archiveprefix = {arXiv},
	Author = {Linch, III, William D. and McOrist, Jock and Vallilo, Brenno Carlini},
	Doi = {10.1088/1126-6708/2008/09/042},
	Eprint = {0804.0613},
	Journal = {JHEP},
	Pages = {042},
	Primaryclass = {hep-th},
	Reportnumber = {EFI-08-08, YITP-SB-08-13},
	Slaccitation = {%%CITATION = ARXIV:0804.0613;%%},
	Title = {{Type IIB Flux Vacua from the String Worldsheet}},
	Volume = {09},
	Year = {2008}}

@article{Chandia:2003hn,
	Archiveprefix = {arXiv},
	Author = {Chandia, Osvaldo and Vallilo, Brenno Carlini},
	Doi = {10.1088/1126-6708/2004/04/041},
	Eprint = {hep-th/0401226},
	Journal = {JHEP},
	Pages = {041},
	Primaryclass = {hep-th},
	Reportnumber = {DFPD-04-TH-04},
	Slaccitation = {%%CITATION = HEP-TH/0401226;%%},
	Title = {{Conformal invariance of the pure spinor superstring in a curved background}},
	Volume = {04},
	Year = {2004}}

@article{Berkovits:2014ama,
	Archiveprefix = {arXiv},
	Author = {Berkovits, Nathan and Chandia, Osvaldo},
	Doi = {10.1007/JHEP06(2014)001},
	Eprint = {1403.2429},
	Journal = {JHEP},
	Pages = {001},
	Primaryclass = {hep-th},
	Slaccitation = {%%CITATION = ARXIV:1403.2429;%%},
	Title = {{Simplified Pure Spinor b Ghost in a Curved Heterotic Superstring Background}},
	Volume = {06},
	Year = {2014}}

@article{Berkovits:2001ue,
	Archiveprefix = {arXiv},
	Author = {Berkovits, Nathan and Howe, Paul S.},
	Doi = {10.1016/S0550-3213(02)00352-8},
	Eprint = {hep-th/0112160},
	Journal = {Nucl. Phys.},
	Pages = {75-105},
	Primaryclass = {hep-th},
	Reportnumber = {IFT-P-072-2001, KCL-TH-01-49, IFT-P.072-2001},
	Slaccitation = {%%CITATION = HEP-TH/0112160;%%},
	Title = {{Ten-dimensional supergravity constraints from the pure spinor formalism for the superstring}},
	Volume = {B635},
	Year = {2002}}

@article{Berkovits:2000fe,
	Archiveprefix = {arXiv},
	Author = {Berkovits, Nathan},
	Doi = {10.1088/1126-6708/2000/04/018},
	Eprint = {hep-th/0001035},
	Journal = {JHEP},
	Pages = {018},
	Primaryclass = {hep-th},
	Reportnumber = {IFT-P-005-2000},
	Slaccitation = {%%CITATION = HEP-TH/0001035;%%},
	Title = {{Super Poincare covariant quantization of the superstring}},
	Volume = {04},
	Year = {2000}}

@article{Berkovits:1994wr,
	Archiveprefix = {arXiv},
	Author = {Berkovits, Nathan},
	Doi = {10.1016/0550-3213(94)90106-6},
	Eprint = {hep-th/9404162},
	Journal = {Nucl. Phys.},
	Pages = {258-272},
	Primaryclass = {hep-th},
	Reportnumber = {KCL-TH-94-5},
	Slaccitation = {%%CITATION = HEP-TH/9404162;%%},
	Title = {{Covariant quantization of the Green-Schwarz superstring in a Calabi-Yau background}},
	Volume = {B431},
	Year = {1994}}

@article{Chandia:2011wd,
	Archiveprefix = {arXiv},
	Author = {Chandia, Osvaldo and Linch, William D. and Carlini Vallilo, Brenno},
	Doi = {10.1007/JHEP10(2011)098},
	Eprint = {1108.3555},
	Journal = {JHEP},
	Pages = {098},
	Primaryclass = {hep-th},
	Slaccitation = {%%CITATION = ARXIV:1108.3555;%%},
	Title = {{Compactification of the Heterotic Pure Spinor Superstring II}},
	Volume = {10},
	Year = {2011}}

@article{Chandia:2009it,
	Archiveprefix = {arXiv},
	Author = {Chandia, Osvaldo and Linch, III, William D. and Vallilo, Brenno Carlini},
	Doi = {10.1088/1126-6708/2009/10/060},
	Eprint = {0907.2247},
	Journal = {JHEP},
	Pages = {060},
	Primaryclass = {hep-th},
	Reportnumber = {YITP-SB-09-21},
	Slaccitation = {%%CITATION = ARXIV:0907.2247;%%},
	Title = {{Compactification of the Heterotic Pure Spinor Superstring I}},
	Volume = {10},
	Year = {2009}}

@article{Linch:2006ig,
	Archiveprefix = {arXiv},
	Author = {Linch, III, William D. and Vallilo, Brenno Carlini},
	Doi = {10.1088/1126-6708/2007/01/099},
	Eprint = {hep-th/0607122},
	Journal = {JHEP},
	Pages = {099},
	Primaryclass = {hep-th},
	Reportnumber = {YITP-SB-06-30},
	Slaccitation = {%%CITATION = HEP-TH/0607122;%%},
	Title = {{Hybrid formalism, supersymmetry reduction, and Ramond-Ramond fluxes}},
	Volume = {01},
	Year = {2007}}

@article{Chandia:2006ix,
	Archiveprefix = {arXiv},
	Author = {Chandia, Osvaldo},
	Doi = {10.1088/1126-6708/2006/07/019},
	Eprint = {hep-th/0604115},
	Journal = {JHEP},
	Pages = {019},
	Primaryclass = {hep-th},
	Slaccitation = {%%CITATION = HEP-TH/0604115;%%},
	Title = {{A Note on the classical BRST symmetry of the pure spinor string in a curved background}},
	Volume = {07},
	Year = {2006}}

@article{Chandia:2021coc,
    author = "Chandia, Osvaldo and Vallilo, Brenno Carlini",
    title = "{Relating the $b$ ghost and the vertex operators of the pure spinor superstring}",
    eprint = "2101.01129",
    archivePrefix = "arXiv",
    primaryClass = "hep-th",
    doi = "10.1007/JHEP03(2021)165",
    journal = "JHEP",
    volume = "03",
    pages = "165",
    year = "2021"
}

@article{Berkovits:2022dbm,
    author = "Berkovits, Nathan and Chandia, Osvaldo and Gomide, Joao and Martins, Lucas N. S.",
    title = "{B-RNS-GSS heterotic string in curved backgrounds}",
    eprint = "2211.06899",
    archivePrefix = "arXiv",
    primaryClass = "hep-th",
    doi = "10.1007/JHEP02(2023)102",
    journal = "JHEP",
    volume = "02",
    pages = "102",
    year = "2023"
}

@article{Chandia:2023eel,
    author = "Chandia, Osvaldo and Gomide, Joao",
    title = "{B-RNS-GSS type II superstring in Ramond-Ramond backgrounds}",
    eprint = "2310.02182",
    archivePrefix = "arXiv",
    primaryClass = "hep-th",
    doi = "10.1007/JHEP01(2024)064",
    journal = "JHEP",
    volume = "01",
    pages = "064",
    year = "2024"
}

@article{Berkovits:2021xwh,
    author = "Berkovits, Nathan",
    title = "{Manifest spacetime supersymmetry and the superstring}",
    eprint = "2106.04448",
    archivePrefix = "arXiv",
    primaryClass = "hep-th",
    doi = "10.1007/JHEP10(2021)162",
    journal = "JHEP",
    volume = "10",
    pages = "162",
    year = "2021"
}

@article{McArthur:1983fm,
    author = "McArthur, Ian N.",
    title = "{Superspace normal coordinates}",
    reportNumber = "HUTP-83/A051",
    doi = "10.1088/0264-9381/1/3/003",
    journal = "Class. Quant. Grav.",
    volume = "1",
    pages = "233",
    year = "1984"
}

@article{Chandia:2025nxx,
    author = "Chandia, Osvaldo",
    title = "{A note on type II superstring vertex operators in the B-RNS-GSS formalism}",
    eprint = "2507.05492",
    archivePrefix = "arXiv",
    primaryClass = "hep-th",
    doi = "10.1140/epjc/s10052-025-15045-5",
    journal = "Eur. Phys. J. C",
    volume = "85",
    number = "11",
    pages = "1287",
    year = "2025"
}

@article{Chandia:2013ima,
    author = "Chandia, Osvaldo",
    title = "{The Non-minimal Heterotic Pure Spinor String in a Curved Background}",
    eprint = "1311.7012",
    archivePrefix = "arXiv",
    primaryClass = "hep-th",
    doi = "10.1007/JHEP03(2014)095",
    journal = "JHEP",
    volume = "03",
    pages = "095",
    year = "2014"
}
}
\end{document}